\documentclass{ws-rv9x6}
\usepackage{subfigure}   
\usepackage{ws-rv-thm}   
\usepackage{ws-rv-van}   
\makeindex

\newcommand{\mr}{\mathrm}

\newcommand{\lep}{$l^+l^-$}
\newcommand{\WW}{$W^+W^-$}
\newcommand{\ggll}{$\gamma\gamma\rightarrow l^+l^-$}
\newcommand{\ggWW}{$\gamma\gamma\rightarrow W^+W^-$}
\newcommand{\WWemX}{$W^+W^-\rightarrow e^{\pm}\mu^{\mp}X$}
\newcommand{\xit}{$\tilde{\xi}$}
\begin{document}
\chapter{ATLAS results on diffraction and exclusive production}

\author{Marek Ta\v{s}evsk\'{y}\footnote{Marek.Tasevsky@cern.ch} \\
on behalf of the ATLAS Collaboration}

\address{Institute of Physics of the Academy of sciences of the Czech Republic,\\
Na Slovance 2, 18221 Prague, Czech Republic
}





\begin{abstract}
Various aspects of forward physics have been studied by the ATLAS collaboration 
using data from Run I at the LHC. In this text, main results of three published 
analyses are summarized, based on data from proton-proton collisions at 
$\sqrt{s} = 7$ or 8~TeV collected between 2010 and 2012. One analysis deals with
diffractive signature with at least two jets in the final state, the other two
study exclusive production of a pair of leptons or W bosons.
\end{abstract}



\section{Introduction}
A better understanding of diffraction and exclusive processes at LHC is not
only useful on its own (an experimentalist's view on the diffraction physics
program and early exclusive measurements at LHC is discussed in \cite{MTDiffr}
and \cite{MTexcl}, respectively, it is also profitable for other LHC analyses
where
they form a non-negligible background. Both types of processes were measured at
HERA and Tevatron but cross sections are still known with a limited precision
at LHC. Furthermore, results of such measurements are important in various
Monte Carlo tunes. 
All three presented processes have a common feature regarding the final state,
namely the existence of rapidity gaps, i.e. regions in detector devoid of
hadronic activity. The large rapidity gaps, or intact
proton in the final state, refer to the colorless exchange. Since the intact
proton was not measured in the data used in these analyses (the special forward
proton detectors around the ATLAS detector \cite{ATLASTDR} were only installed
during the Run II), we are left with two approaches to select the signal and
reduce the background. Either we select only events with large
rapidity gaps in the central detector and then we are forced to concentrate on
low pile-up and measure processes with large cross-sections; or we require no
tracks and vertices around the lepton vertex and then we can also use data
samples with a large pile-up and hence access processes with low cross-sections.
The former approach was used in the analysis of diffractive dijets where used
data come from first runs of LHC with a relatively low pile-up. The latter
approach was then used in the remaining two analyses where much larger data
samples were analyzed but also with a more significant amount of pile-up. 

\begin{figure}
  \includegraphics[height=.17\textheight,width=0.28\textwidth]{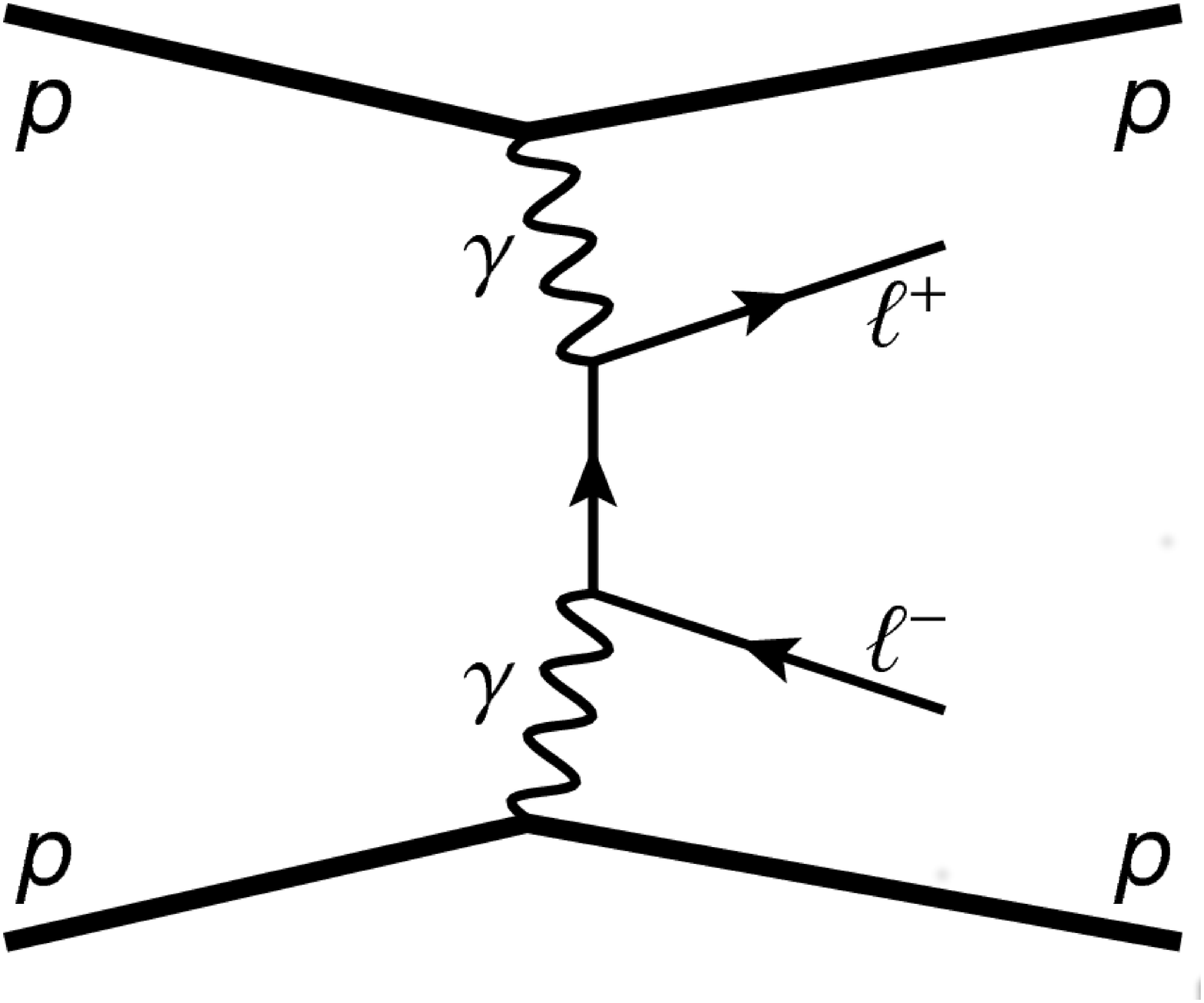}
  \hspace*{0.6cm}
  \includegraphics[height=.17\textheight,width=0.28\textwidth]{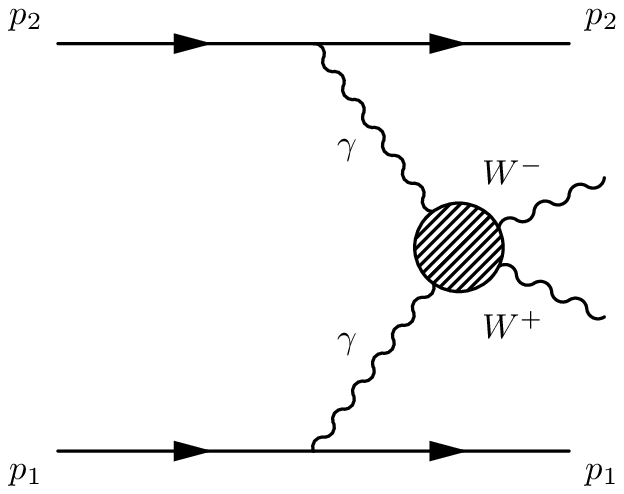}
  \hspace*{0.6cm}
\includegraphics[height=.17\textheight,width=0.28\textwidth]{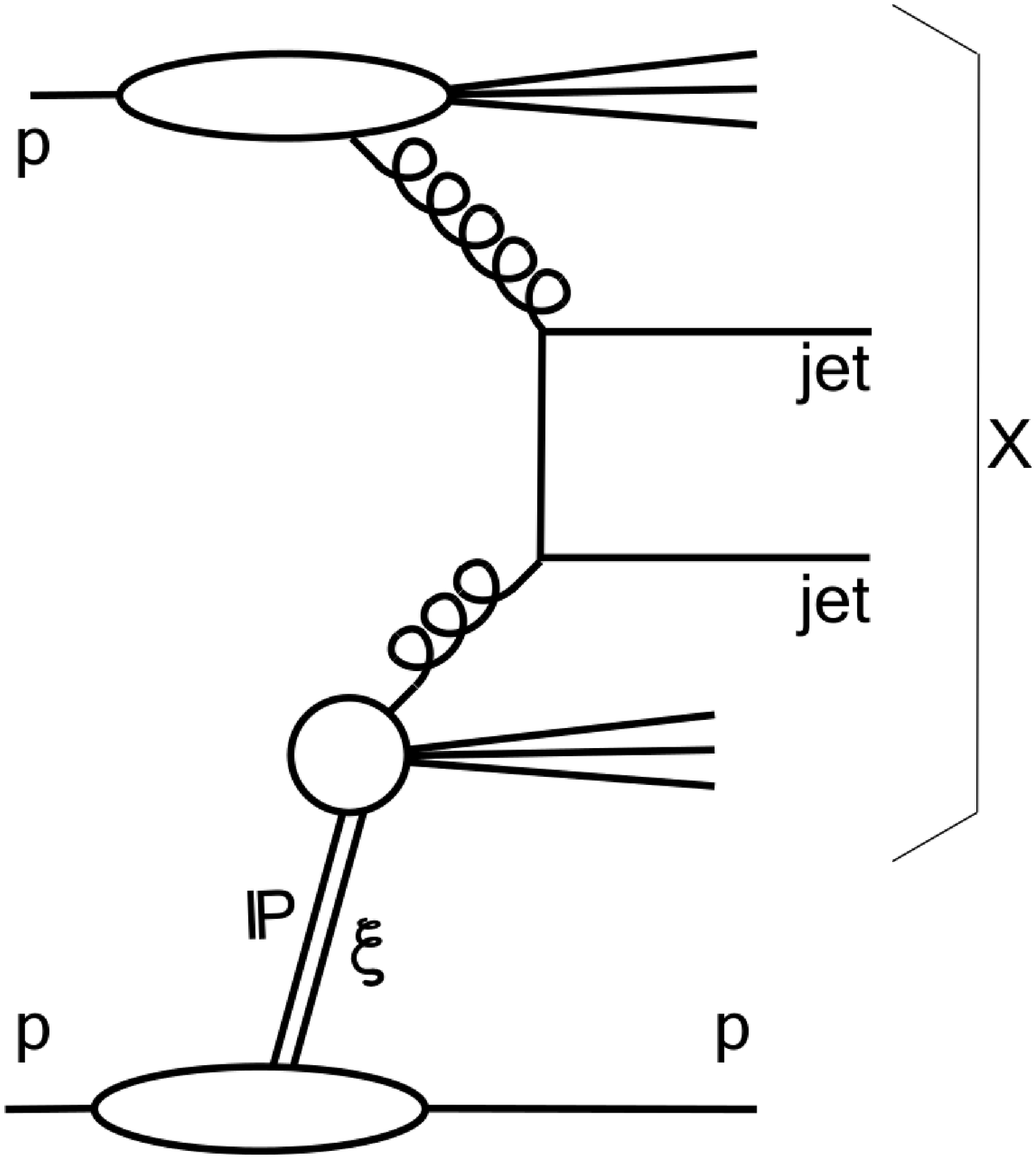}
\caption{Feynman diagrams for the signal processes of analyses reported in
this text: left) QED \ggll, middle) QED \ggWW\ and right) SD dijets.}
\label{diagrams}
\end{figure}

The processes with intact protons are divided into two broad classes. The
photon-induced processes which are calculable using Quantum Electrodynamics
(QED) include QED diffractive processes such as Single dissociation (SD), Double
dissociation (DD) and QED exclusive processes, which can be calculated with a
precision of down to 2\% (based on the Equivalent photon approximation (EPA))
but where proton absorptive corrections can reach up to 20\%. The QED exclusive
processes with \lep\ or \WW\ in the final state are the signal in two analyses
reported in this text and their Feynman diagrams are shown in
Fig.~\ref{diagrams} left and middle.
  
The second class of processes is formed by parton-induced processes calculable
via Quantum Chromodynamics (QCD) and including SD, DD and Double Pomeron
Exchange (DPE) processes for which we also need to consider the so called soft
survival probabilities if we study them at hadron colliders. The QCD exclusive
processes are not discussed in this text. The process of interest in the
remaining analysis reported here is dijet production in SD events, see
Fig.~\ref{diagrams} right. 
\vspace*{-0.5cm}  
\section{Diffractive dijets}
\begin{figure}
  \includegraphics[height=.3\textheight,width=0.49\textwidth]{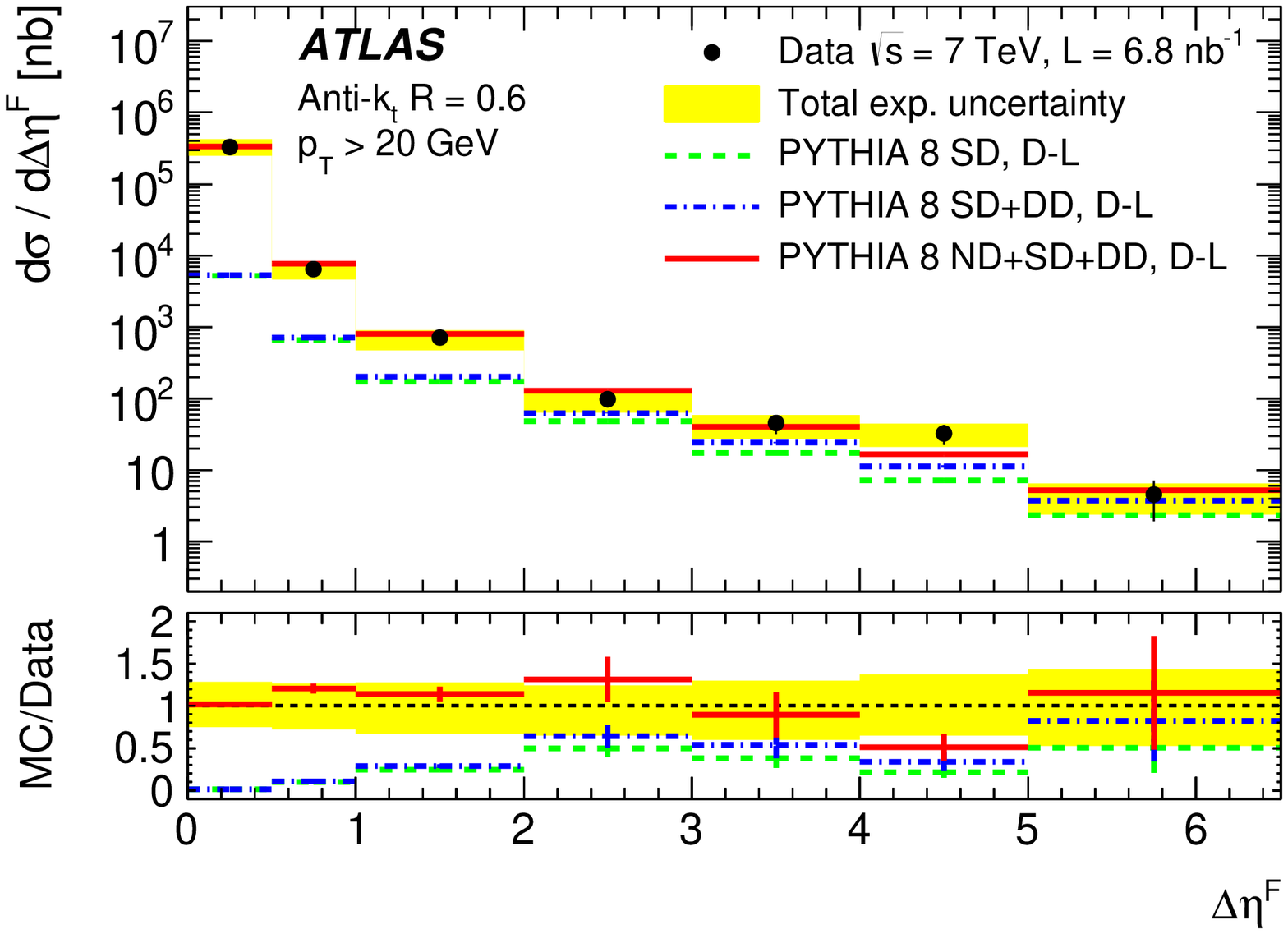}
  \includegraphics[height=.3\textheight,width=0.49\textwidth]{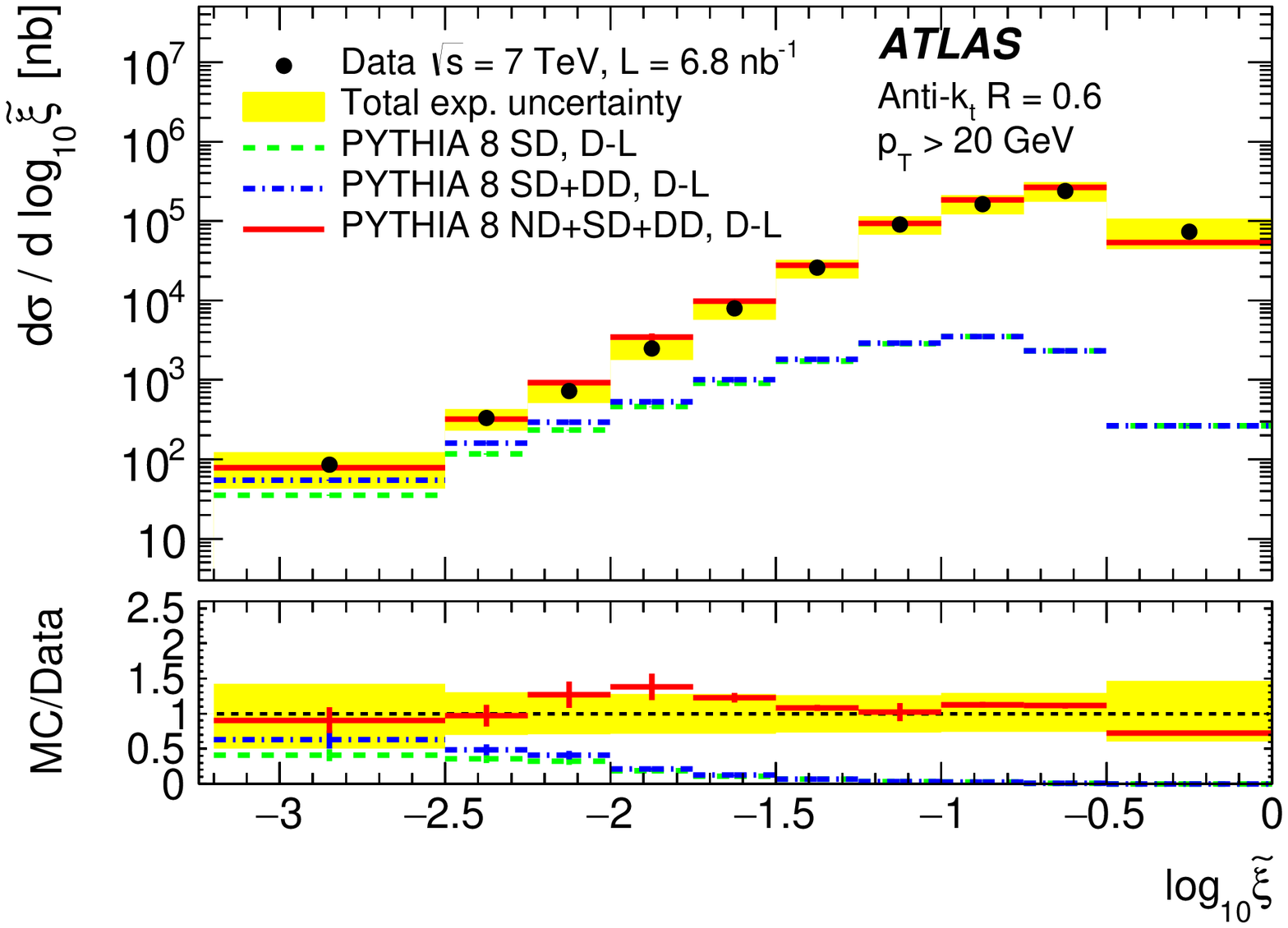}
 \hspace*{-0.2cm}
  \caption{The differential dijet cross sections left) $\Delta\eta_F$ and right)
$\tilde{\xi}$, compared with the particle-level PYTHIA8 model of the SD, sum of
    SD and DD, and sum of all three ND, SD and DD components. Taken from
    Ref.~\cite{ddijets}.}
\label{Djets}
\end{figure}

\vspace*{-0.1cm}
Dijet events with large rapidity gaps between the intact proton (or the edge
of the central detector) and the dijet system were already studied at HERA and
Tevatron. In the CDF measurement \cite{CDFfb} a huge factor of discrepancy
($\sim$10) between the
measured data and theory predictions based on diffractive parton density
functions (dPDF) measured at HERA was observed. This discrepancy translates into
the fact that measured dPDFs are not universal, in other words that the
factorisation is broken. This is usually explained by rescattering of the
dissociated system with the intact proton and the amount of this discrepancy is
often called rapidity gap survival probablity or simply soft survival
probability, $S^2$. In most of cases at HERA the factorisation holds and hence
$S^2 \approx 1.0$, since it is the photon, a much simpler system than the
proton, that enters the diffraction reaction and hence no other rescattering is
expected. Nevertheless also at HERA the factorisation may be broken in special
cases where this photon has time to develop its structure and hence to be
resolved \cite{HERAfb}. A natural question thus is what value of $S^2$ we can
expect for the
same process, dijets with rapidity gap, at LHC. It should be, however, stressed
that the concept of $S^2$ is not well-defined theoretically. In some models
this probability is embedded in amplitude calculations, in other models it is
possible to factorise it. In any case, this quantity depends on the process
studied and on kinematics. In \cite{ddijets} it was obtained by
comparing the data to the Pomwig model after subtracting from data the
background
from non-diffractive processes (ND) and DD using Pythia 8.1. The measured value
is $S^2 = 16\pm4\pm8$~\% with
rather large systematic uncertainties coming from model dependence.
Results that are easier to interpret are corrected cross-sections as functions
of the forward gap size, $\Delta\eta_F$ and the fraction of
four-momentum of the incident proton carried by Pomeron, \xit. From
Fig.~\ref{Djets} we can see that the gap plateau observed and expected in the
gap analysis with no jet requirement \cite{gaps} is not observed here, which can
be simply
explained by the topology: the presence of at least two jets, each with a radius
$R=0.6$ does not leave much space for large gaps.
Fig.~\ref{Djets} also provides a comparison of the data with Pythia 8.1 where
the ND contribution is normalized to match the data in the first
\xit\ bin and the SD and DD contributions are used with default
cross-sections.
The striking feature of Pythia 8.1 model is that the sum of ND, SD and DD
contributions describes the data satisfactorily, i.e. with $S^2 \approx 1.0$.
We also see that the ND contribution extends to fairly large gaps and small
$\tilde{\xi}$ values.

\section{Exclusive dileptons}\label{El}
\begin{figure}
\begin{center}
\includegraphics[height=.3\textheight,width=0.6\textwidth]{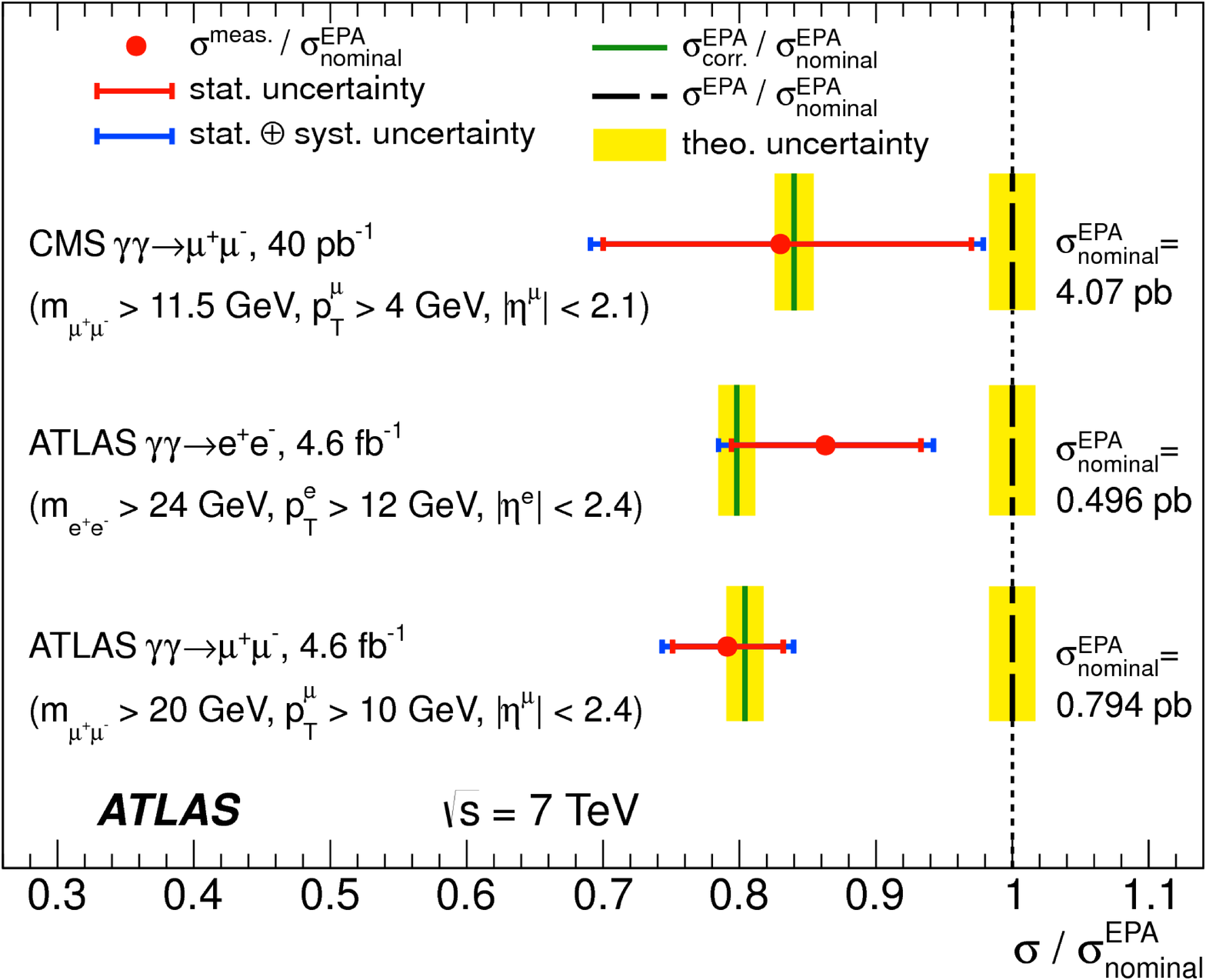}
\caption{Comparison of the ratios of measured (red points) and predicted (solid
  green lines) cross-sections to the uncorrected EPA calculations (black dashed
  line). Taken from Ref.~\cite{exclll}.}
  \label{excllep}
  \end{center}
\end{figure}

In analysis \cite{exclll} the selection focused on electron and muon pairs
produced exclusively,
i.e. with no other activity around the production vertex in the central
detector. This is a standard candle process thanks to its simple final state
which can also be used for a luminosity calibration as well as for alignment and
calibration of forward proton detectors AFP \cite{AFP} and CT-PPS \cite{CT-PPS}.
The cross-sections, however, are small and hence only calibration/alignment
over larger data taking periods would be possible. A key requirement is the so
called exclusivity veto demanding no tracks with transverse momentum
$p_T > 0.4$~GeV from the lepton vertex and in addition no tracks and vertices
within at least 3~mm from the longitudinal isolation of the lepton vertex. Then
after requiring the dilepton mass to be in a window of 70 and 105~GeV and
restricting the lepton $p_T$ to be below 1.5~GeV, two acoplanarity distributions
are made, namely $1-|\Delta\phi_{e^+e^-}|/\pi$ and $1-|\Delta\phi_{\mu^+\mu^-}|/\pi$
where data are compared to predictions of Herwig++ for exclusive, of LPAIR for
SD and of Powheg and Pythia for DD and Drell Yan processes. The best description
of the data is reached when the exclusive and SD contributions are scaled by
factors 0.863 and 0.759, respectively, for the $e^+e^-$ pair, or by 0.791 and
0.762, respectively, for the muon pair. It is important to note that these
scaling factors were found in agreement with those predicted in
Ref.~\cite{HLMR1}.
The measured fiducial cross-sections for the exclusive \ggll\ are $\sigma_{\mr excl}(\gamma\gamma\rightarrow e^+e^-) = 0.428 \pm 0.035$ (stat) $\pm 0.018$ (syst) pb and $\sigma_{\mr excl}(\gamma\gamma\rightarrow \mu^+\mu^-) = 0.628 \pm 0.032$ (stat) $\pm 0.021$ (syst) pb. These values are found
in a very good agreement with cross-sections based on EPA and corrected for
absorptive corrections which range around 20\% as well as with CMS measurement
(see Fig.~\ref{excllep}).

\section{Exclusive W bosons}\label{EW}
\begin{figure}
  \includegraphics[height=.3\textheight,width=0.49\textwidth]{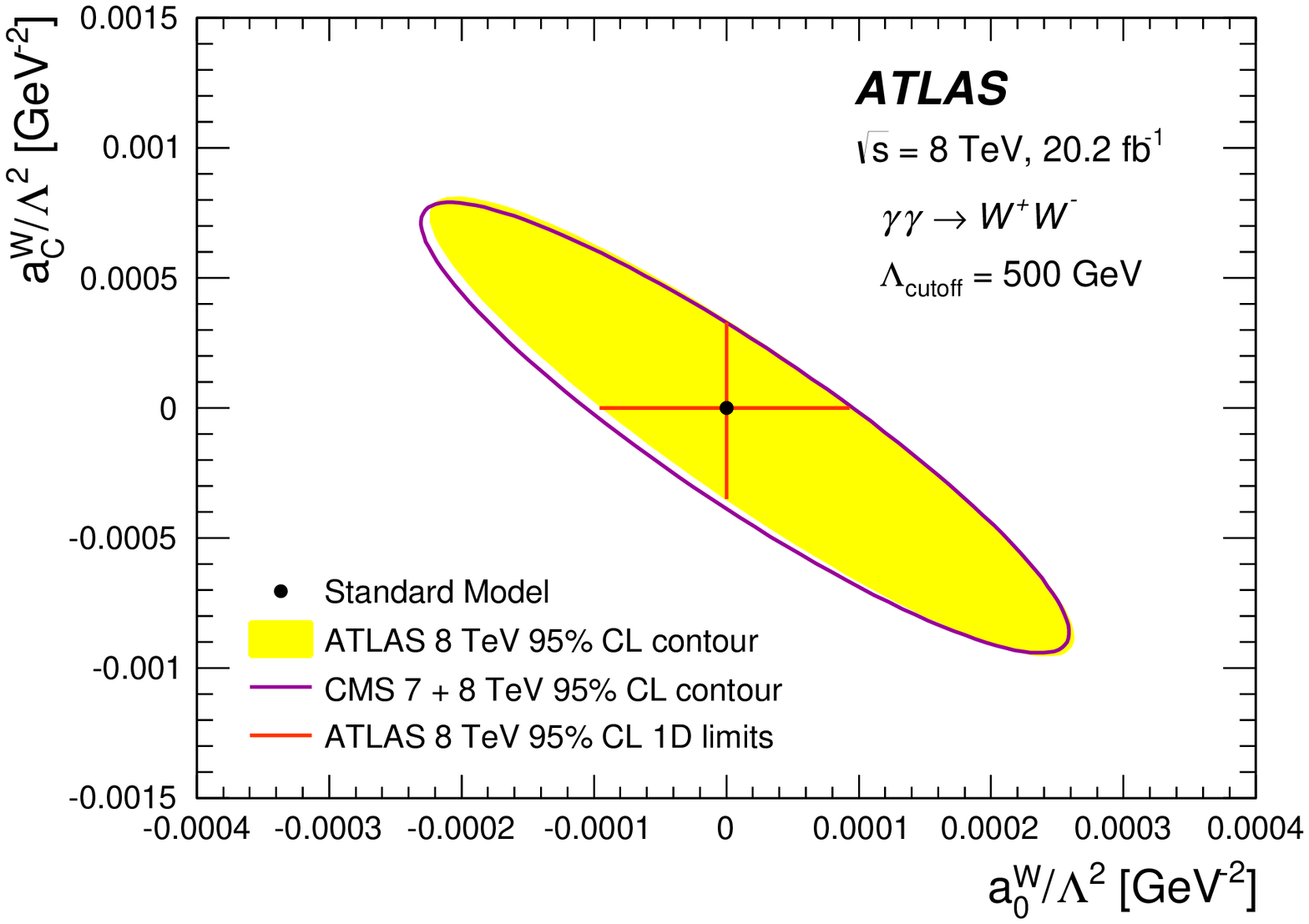}
  \includegraphics[height=.3\textheight,width=0.49\textwidth]{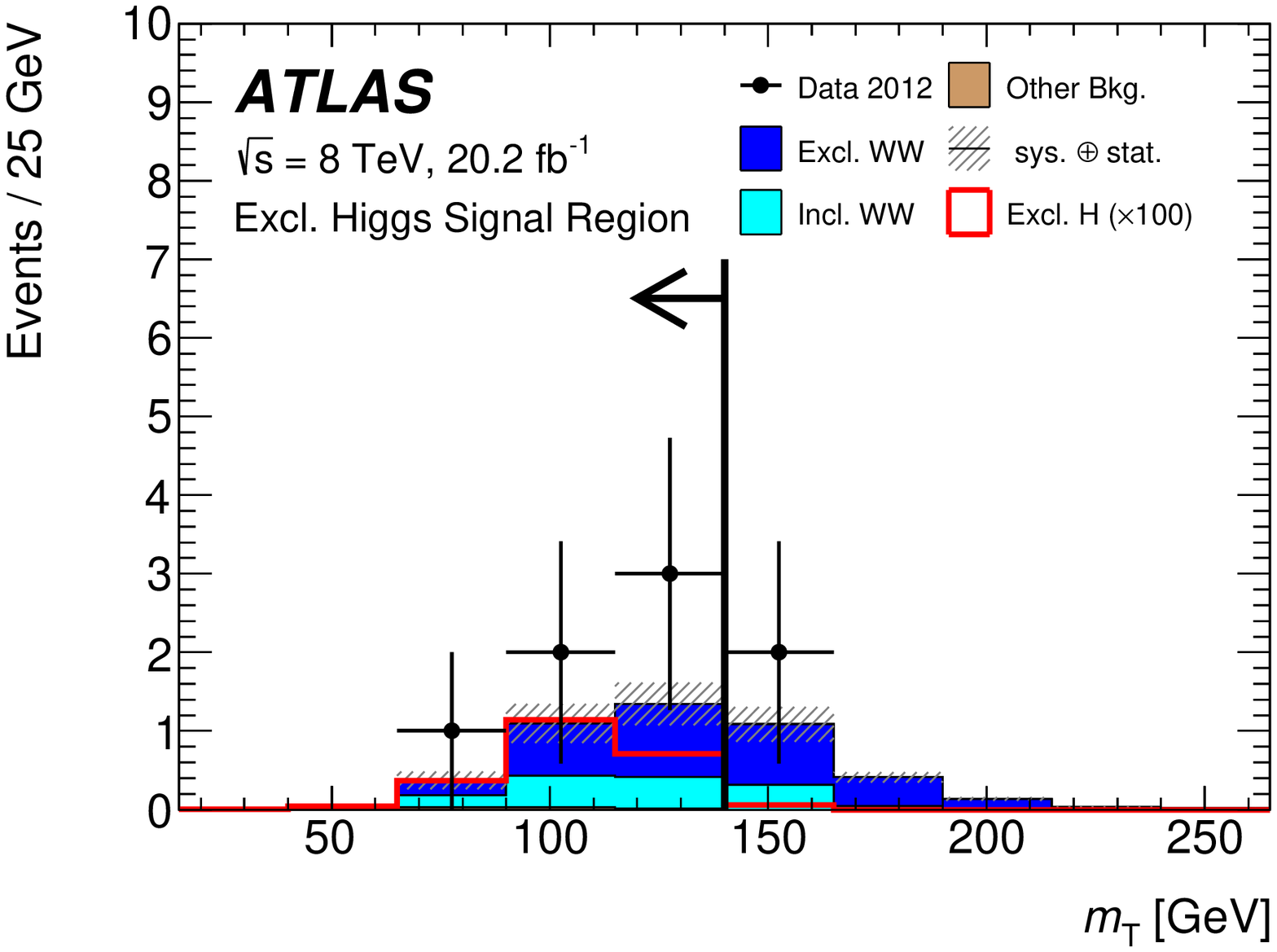}
  \caption{left) The observed log-likelihood 95\% confidence-level contour and
    1D limits for the case with a dipole form factor with $\Lambda_{\mr cutoff} = 500$~GeV, right) distribution of transverse mass in the signal region of exclusive
Higgs. The arrow denotes the selection. Taken from Ref.~\cite{exclWW}.}
\label{exclWW}
\end{figure}

The analysis described in Ref.~\cite{exclWW} aims at estimating anomalous
quartic gauge couplings (aQGC) $\gamma\gamma\,WW$ and collecting first
candidates of exclusive Higgs boson decaying into the \WW\ pair decaying
further to $e^{\pm}\nu\mu^{\mp}\nu$. This
analysis profits from the measurement of the exclusive leptons presented in the
previous section. Again the key requirement is the exclusivity veto described
above, only the dilepton vertex longitudinal isolation is required to be less
than 1~mm because in this sample data are more contaminated by pile-up.
Similarly also the acomplanarity distributions are
constructed, from which a ratio of observed to predicted QED
exclusive events is extracted to be 0.76 which is in a reasonable agreement with
values obtained in the dilepton analysis. Non-existent simulation of
SD and DD \ggWW\ processes is accounted for by multiplying
predicted QED exclusive \ggWW\ events by a factor 3.3 obtained using exclusive
\ggll\ events in the region $m_{ll} >$160~GeV from Herwig++. This number is found to be in
agreement with that predicted in Ref.~\cite{HLMR2}. The measured cross-section for
the process \ggWW\ in Standard Model extrapolated to the full \WWemX\ phase space is
$\sigma_{\mr excl}($\ggWW$) = 6.9 \pm 2.2$ (stat) $\pm$ 1.4 (syst)~fb which can be
compared to the predicted Herwig++ cross section of 4.4 $\pm$ 0.3~fb. The
background-only hypothesis corresponds to a significance of 3.0. Limits on
aQGC are obtained using event yields in the distribution of $p_T$ of the
electron-muon pair for $p_T^{e\mu} > 120$~GeV and are visualised in
Fig.~\ref{exclWW}. We can see that the limits are compatible with those by CMS
and they are more stringent than those published by OPAL, D0 and CMS.
While studied in detail phenomonelogically (see e.g. \cite{diffH1,HWW,diffH3}),
this analysis reports on finding first exclusive Higgs boson candidates
at LHC experimentally.  
Event yields for the exclusive H$\rightarrow$\WWemX\ were obtained from
Fig.~\ref{exclWW} right in the cut region and they amount to
6 for data, 0.023 $\pm$ 0.003 for signal and 3.0 $\pm$ 0.8 for background. The
signal is obtained using KMR calculations which are based on gluon-induced
production \cite{KMR} and the background is dominantly the exclusive \WW\ and
inclusive \WW\ processes. These yields are then converted to the exclusive Higgs
boson total production cross-section using the CLs technique \cite{CLs} which
gives $\sigma < 1.2$~pb at 95\% CL (observed) and $\sigma < 0.7$~pb at 95\% CL
(expected). Since the cross-section for the exclusive Higgs production by KMR
is about 3~fb, the observed upper limit is 400 times higher. 

\section{Summary}
In the analysis of diffractive dijets, ATLAS collaboration has measured rapidity
gap and \xit\ distributions in the presence of a hard scale. No gap plateau is
observed and comparisons with predictions by Pythia 8.1 show that ND processes
extend to very large gaps and small \xit\ which eventually translates to no
need for gap survival factor to describe the data. In the measurement of
exclusive (photon-induced) events with the dilepton in the final state we
confirm the necessity of absorptive corrections (20\%) to calculations based on
EPA. In the \WW\ final state we obtained an evidence for the
\ggWW\ process, we improved limits on aQGC but have
observed no excess and we also set first observed upper limits for the total
cross-section of the exclusive Higgs production.

\section{Acknowledgments}
Supported by the project LG15052 and LM2015058 of the Ministry of Education of
the Czech Republic.

\bibliographystyle{ws-rv-van}


\begin{thebibliography}{99}
  \footnotesize{
\bibitem{MTDiffr} M.~Tasevsky, {\it Diffractive physics program in ATLAS
  experiment}, Nucl. Phys. Proc. Suppl. {\bf 179-180} (2008) 187;
  ATL-PHYS-CONF-2008-019.
\bibitem{MTexcl} M.~Tasevsky, {\it Measuring Central Exclusive Processes at
  LHC}, arXiv:0910.5205 [hep-ph]. 
\bibitem{ATLASTDR} ATLAS Collab., {\it The ATLAS Experiment at the CERN Large Hadron Collider}, J. Inst. {\bf 3} (2008) S08003.
\bibitem{CDFfb} CDF Collab., {\it Diffractive Dijets with a Leading Antiproton in $\bar{\mr p}p$ Collisions at $\sqrt{s}=1800$~GeV},
  Phys. Rev. Lett. {\bf 84}, 5083 (2000).
\bibitem{HERAfb} HERA Collab., {\it Diffractive Dijet Production with a Leading
  Proton in ep Collisions at HERA}, JHEP {\bf 1505} (2015) 056. 
\bibitem{ddijets} ATLAS Collab., {\it Dijet production in $\sqrt{s}=7$~TeV
pp collisions with large rapidity gaps at the ATLAS experiment},
  Phys. Lett. {\bf B 754} (2016) 214.
\bibitem{gaps} ATLAS Collab., {\it Rapidity gap cross sections measured with the ATLAS detector in pp collisions at $\sqrt{s}=7$~TeV},
  Eur. Phys. J. {\bf C72} (2012) 1926.
\bibitem{AFP} ATLAS  Collab., Technical Design Report for the ATLAS
  Forward Proton Detector,~CERN-LHCC-2015-009, 2015, url: http://cds.cern.ch/record/2017378/; M. Tasevsky, {\it Status of the AFP project in the ATLAS experiment}, AIP Conf. Proc. 1654 (2015) 090001.
  \bibitem{CT-PPS} M. Albrow et al., {\it CMS-TOTEM Precision Proton Spectrometer}, CERN-LHCC-2014-021 ; TOTEM-TDR-003 ; CMS-TDR-13.
\bibitem{exclll} ATLAS Collab., {\it Measurement of exclusive \ggll\
  production in proton–proton collisions at $\sqrt{s}$=7~TeV with the ATLAS
  detector},
  Phys. Lett. {\bf B 749} (2015) 242.
\bibitem{HLMR1} L. Harlan-Lang, V. Khoze and M. Ryskin, {\it Exclusive physics at the LHC with SuperChic 2}, Eur.Phys. J. {\bf C76} nr.1 (2016) 9.
\bibitem{exclWW} ATLAS Collab., {\it Measurement of exclusive \ggWW\
production and search for exclusive Higgs boson production in pp collisions at
$\sqrt{s}=8$~TeV using the ATLAS detector}, Phys. Rev. {\bf D94} (2016) 032011.
\bibitem{HLMR2} L. Harlan-Lang, V. Khoze and M. Ryskin, {\it The photon PDF in events with rapidity gaps}, Eur.Phys. J. {\bf C76} nr.5 (2016) 255.
\bibitem{diffH1} S. Heinemeyer et al., {\it Studying the MSSM Higgs sector by
  forward proton tagging at the LHC}, Eur. Phys. J. {\bf C53} (2008) 231. 
\bibitem{HWW} B. Cox et al., {\it Detecting the standard model Higgs boson in
  the WW decay channel using forward proton tagging at the LHC},
  Eur. Phys. J. {\bf C45} (2006) 401.
\bibitem{diffH3} M. Tasevsky, {\it Exclusive MSSM Higgs production at the LHC after Run I}, Eur.Phys.J. {\bf C73} (2013) 2672.  
\bibitem{KMR} V. Khoze, A. Martin and M. Ryskin, {\it Prospects for new physics observations in diffractive processes at the LHC and Tevatron},
  Eur. Phys. J. {\bf C23} (2002) 311.
\bibitem{CLs}  A. L. Read, {\it Presentation of search results: The CL(s)
  technique}, J. Phys. {\bf G28} (2002) 2693.}
\end{thebibliography}

\end{document}